# AutoMeKin2021: An open-source program for automated reaction discovery

Emilio Martínez-Núñez,[1] George L. Barnes,[2] David R. Glowacki,[3] Sabine Kopec,[4] Daniel Peláez,[4] Aurelio Rodríguez,[5] Roberto Rodríguez-Fernández,[1] Robin J. Shannon,[3] James J. P. Stewart,[6] Pablo G. Tahoces[7] and Saulo A. Vazquez[1]

Correspondence to: Emilio Martínez-Núñez (E-mail: emilio.nunez@usc.es)

[1]*Department of Physical Chemistry, University of Santiago de Compostela, 15782 Santiago de Compostela, Spain*
[2]*Department of Chemistry and Biochemistry, Siena College, 515 Loudon Road, Loudonville, NY, United States*
[3]*Centre for Computational Chemistry, School of Chemistry, University of Bristol, Cantock's Close, Bristol BS8 1TS, UK*
[4]*Institut de Sciences Moléculaires d'Orsay, UMR 8214, Université Paris-Sud - Université Paris-Saclay, 91405 Orsay, France*
[5]*Galicia Supercomputing Center (CESGA), Santiago de Compostela, Spain*
[6]*Stewart Computational Chemistry, 15210 Paddington Circle, Colorado Springs, CO 80921 USA*
[7]*Department of Electronics and Computer Science, University of Santiago de Compostela, 15782 Santiago de Compostela, Spain*

## ABSTRACT

AutoMeKin2021 is an updated version of tsscds2018, a program for the automated discovery of reaction mechanisms (*J. Comput. Chem.* **2018**, *39*, 1922-1930). This release features a number of new capabilities: rare-event molecular dynamics simulations to enhance reaction discovery, extension of the original search algorithm to study van der Waals complexes, use of chemical knowledge, a new search algorithm based on bond-order time series analysis, statistics of the chemical reaction networks, a web application to submit jobs, and other features. The source code, manual, installation instructions and the website link are available at: https://rxnkin.usc.es/index.php/AutoMeKin

## Introduction

Over the last several years, computational chemistry has witnessed a surge in the development of methods for reaction mechanism discovery.[1-65] Many of these methods predict complex reaction networks in an automated manner, where the search of reactions is usually more thorough than the traditional "by hand" approach.

Our group is actively involved in this endeavor, and a few years ago we presented a new automated method called Transition State Search using Chemical Dynamics Simulations (TSSCDS).[44,45] Our algorithm relied on a Molecular Dynamics (MD)-based exploration of configurational space, followed by a post-processing analysis to locate promising transition state (TS) candidates from the MD snapshots.[43-45] While other methods also use the ability of MD simulations to discover reaction mechanisms, the distinctive feature of our approach is the focus on finding saddle-point structures. Up until now, locating TSs from MD simulations has been difficult, but the procedure described here has proven to be very effective and useful in predicting unexpected mechanisms. Our approach has been applied in combustion chemistry,[66,67] cycloaddition reactions,[68] photodissociations,[69-71] organometallic catalysis,[43] radiation damage of biological systems,[72] simulation of mass spectrometry experiments,[73,74] and other applications.[75-77]

The first version of the computer program implementing our approach was released three years ago under the name tsscds2018.[46] This approach, along with the algorithms described in this paper, has now been combined and implemented in the software package named AutoMeKin,[78] which stands for **Auto**mated **Me**chanisms and **Kin**etics.

AutoMeKin2021 includes the following new features: (a) the rare-event acceleration method BXDE;[76,79] (b) a generalization of a Graph Theory-based algorithm to locate TS structures for the study of non-covalent interactions;[80] (c) a chemical knowledge-based method for reaction discovery; (d) a new TS search algorithm based on a bond-order time series analysis;[81] (e) a statistical analysis of the chemical reaction networks using the Python library NetworkX;[82] (f) a web application for online job submission; as well as other features.

After a brief introduction of the original method, a description will be given of the new methods incorporated into AutoMeKin2021, as well as some test cases and sample input files. Its new capabilities, as well as some proposed future improvements, will be summarized in the conclusions.

## Methods

AutoMeKin's main components are:

a) Short-time reactive MD simulations
b) Post-processing analysis of the MD simulations
c) Kinetics simulations

To run the MD simulations, a sizeable amount of vibrational energy is adaptatively placed in each vibrational mode, to trigger reactive events. Also, as described below, a new rare-event acceleration technique is available, which allows for an efficient sampling of reactive events by imposing a bias on the potential energy.

In the analysis of the MD trajectories, some concepts from Graph Theory are useful, including the Adjacency and Laplacian matrices and the SPRINT coordinates.[45] These are used in locating suitable TS guesses and in constructing the chemical reaction network. Specifically, the MD snapshots are screened to find TS candidates associated with a reactive event. This is accomplished by transforming a 3D molecular geometry into a graph, which is defined by its adjacency matrix **A**, whose elements, $a_{ij}$, are given by:[45]

$$a_{ij} = \begin{cases} 1 & \text{if } \delta_{ij} < 1 \\ 0 & \text{otherwise} \end{cases}, \text{ with } \delta_{ij} = \frac{r_{ij}}{r_{ij}^{\text{ref}}}, \quad (1)$$

where $r_{ij}$ and $r_{ij}^{\text{ref}}$ are the interatomic and reference distances, respectively, of each pair, $ij$, of atoms. Reference distances are determined from the sum of the covalent radii of atoms $i$ and $j$.

A reactive event is then deemed to occur when for any atom $j$:[45]

$$\max(\delta_{jk}) > \min(\delta_{jl}), \quad (2)$$

where index $k$ runs over the set of atoms that are covalently bonded to $j$ (neighbors), and index $l$ runs over the remaining (non-neighbor) atoms. In other words, the criterion of Eq. 2 is met when the nearest atom to $j$ is a non-neighbor. Since more than two bonds can be part of the reaction coordinate in a given transition state, reactive events occurring within an adaptive time window of ~10-20 fs are merged.[45] The resulting structures are first subjected to a partial relaxation, with the atoms involved in the reactive event kept frozen, and then optimized to a TS (saddle point of index one). This search algorithm is named *bbfs*, which stands for bond breaking/formation search.

For the sake of efficiency, the trajectories are integrated with either MOPAC2016[83] or Entos Qcore[84] at a semiempirical quantum-mechanical (SQM) level, while the stationary points are re-optimized with Gaussian09[85] or Entos Qcore[84] using a higher level of electronic structure theory. More details about the method can be found in the original papers.[44,45]

Table 1 shows a summary of the most important tools that have been implemented in the last version of AutoMeKin. They will be described in the next sections.

**Table 1.** Main tools available in AutoMeKin2021

| Method[a] | Features | Dependencies | Ref |
|---|---|---|---|
| BXDE | Accelerated MD simulation | ASE | 76,79 |
| vdW | Sampling vdW structures | ASE | 80 |
| ChemKnow | Graph transformations and NEB | ASE NetworkX | This work |
| bots | Reactive event search algorithm | - | 81 |
| Reaction network properties | Graph-Theory-based statistics | NetworkX | 76 |
| web application | Online submission of jobs | - | This work |

[a] Name of the tool/method.

**Rare-event acceleration method BXDE**

Standard MD simulations are typically biased towards the entropically-favored reaction pathways. AutoMeKin's standard MD module employs initial conditions with substantial amounts of vibrational energy to accelerate the incidence of reactive events.

An alternative way of accelerating reactive events, called Boxed Molecular Dynamics in Energy space (BXDE)[79] has recently been proposed. BXDE belongs to the family of BXD methods,[86-89] which introduce reflective barriers in the phase space of an MD trajectory along a particular (collective) variable. The boundaries are employed to push the dynamics along the collective variable into regions of phase space which would rarely be sampled in an unbiased trajectory.

In BXDE, the bias is introduced into the potential energy, rather than in any particular collective variable of the system. The different chemical reaction channels are sampled by gradually scanning through potential energy "boxes" or energetic "windows." The BXDE simulation module in AutoMeKin utilizes the Atomistic Simulation Environment (ASE) package, and MOPAC2016 or Entos Qcore are interfaced *via* the ASE calculator class.[90]

```
--Method--
sampling      BXDE
ntraj         1
fs            5000
fric          0.5
post_proc     bbfs 20 1
temp          1000
```

**Figure 1.** Section of an input file for a BXDE calculation.

Figure 1 shows the section of an input file where a BXDE calculation is requested. Each line in the input file consists of a *keyword value* pair. BXDE is one of the different sampling methods employed in AutoMeKin for finding stationary states in a potential energy surface, with other sampling alternatives such as: MD, MD-micro, external, ChemKnow, association, and vdW. Although some of the options are described in this work, the reader is referred to the documentation for more details.[78]

The number of trajectories and simulation time are specified through the keywords *fs* and *ntraj*, respectively. BXDE employs a Langevin thermostat whose friction coefficient in ps$^{-1}$ (keyword *fric*) and temperature in K (keyword *temp*) must be entered as well.

An example of the use of BXDE combined with AutoMeKin's TS search algorithms is the recent study of the ozonolysis of α-pinene.[76] This reaction is known to follow the "Criegee mechanism" of alkene ozonolysis (see Figure 2), and was previously studied using ab initio methods.[91] Despite the low level of electronic structure (PM7) employed to run BXDE and to optimize the stationary points,[76] not only was the new approach capable of predicting the major pathways (shown in black in Figure 2), but a significant number of new intermediates and pathways were also predicted. The figure also shows (in red) some of the most important pathways that were overlooked in the previous ab initio study and found in the BXDE sampling.[76] A full account of the new mechanisms predicted by BXDE is detailed elsewhere.[76]

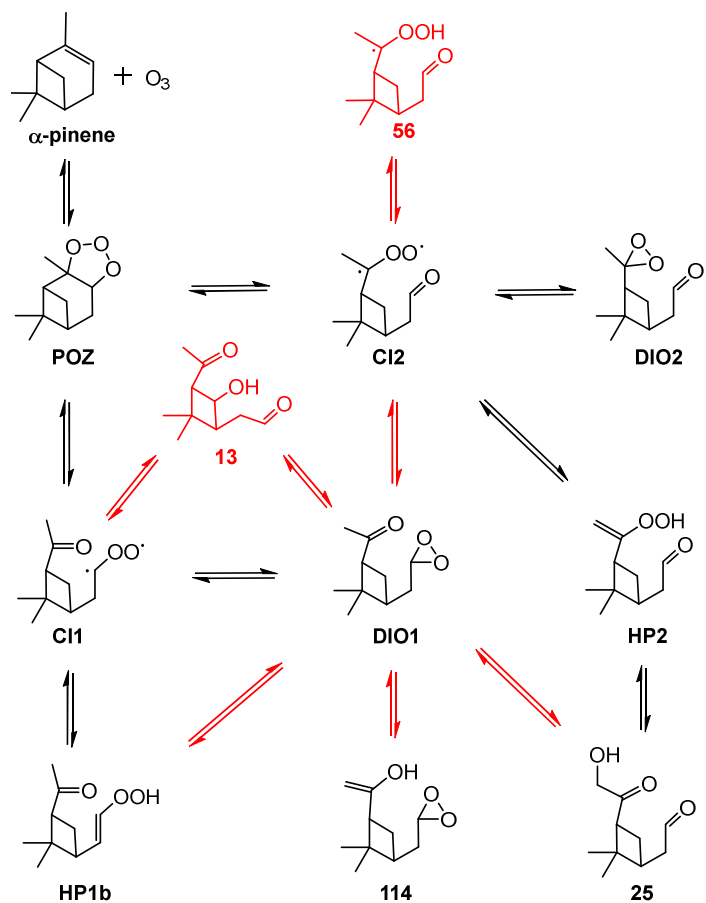

**Figure 2.** Major reaction mechanisms of the
α-pinene ozonolysis, featuring the new reaction pathways found by BXDE[76] (red).

**Non-covalent interactions (vdW)**

The original search algorithm relies on the adjacency matrix of Eq. 1, where the reference distance is determined from the covalent radii of the atoms. Consequently, a complex where two molecules are held together by intermolecular interactions would not be regarded as a single entity, but as two separated fragments.

To expand the scope of the method, it was recently suggested that matrix **A** should be recast in a block structure that accounts for a system made up of molecules B and C:[80]

$$\mathbf{A} = \begin{pmatrix} \mathbf{B} & \mathbf{BC} \\ \mathbf{BC} & \mathbf{C} \end{pmatrix} \quad (3)$$

where the diagonal blocks **B** and **C** refer to the (covalent) connectivity within B and C, respectively, whereas the off-diagonal **BC** block corresponds to the non-covalent, i.e., vdW, interacting system B−C. The matrix elements for **B** and **C** are evaluated according to Eq. 1, with the reference distances determined from the covalent radii. In contrast, the matrix elements of the **BC** block are calculated using Eq. 1 but with the reference distances determined from van der Waals radii.[92]

In this new *ansatz*, non-covalent interactions in B−C are treated on the same footing as covalent ones within any of the fragments, thus permitting the detection of TSs connecting van der Waals (vdW) structures. The method can be easily extended to more than two interacting molecules, and has been recently applied to study the Ar−benzene, $N_2$−benzene, $(H_2O)_n$−benzene ($n$ = 1−3), and $(NO_2$−benzene$)^+$ systems.[80]

To prevent the inherent bias of standard high-energy MD simulations toward dissociation of the complex, the BXDE sampling option is automatically used when a vdW calculation is called for. Figure 3 shows an example input file for a vdW calculation on the pyrene + $NO_2$ system.[93,94]

In this example, a total simulation time of 2 ps is employed. The electronic structure level of theory employed in this example is GFN1-xTB[95] (*xtb*) using the Entos Qcore program,[96] which is requested using the keyword *LowLevel* followed by the computer program (*qcore*) and the method (*xtb*).

To generate the starting structures for the dynamics the keywords *rotate* and *Nassoc* are used. The former has four values:

```
rotate pivotA pivotB r_pivot r_min
```

where *pivotA* and *pivotB* refer to the pivot points for the random rotations of fragments A and B, respectively (center of mass, *com*, for both). Then, *r_pivot* is the fixed distance between the pivot points (4.0 Å), and *r_min* is the minimum distance between any pair of atoms of different fragments (1.5 Å).

The keyword *Nassoc* is used to select the number of initial structures generated. The randomly generated structures are subjected to optimization using *xtb*. The global minimum from the set of optimized structures is then used as the starting point for the MD simulations.

```
--General-
Molecule        pyrene-NO2
fragmentA       pyrene
fragmentB       NO2
LowLevel        qcore xtb

--Method--
sampling        vdW
rotate          com com 4.0 1.5
Nassoc          10
ntraj           1
fs              2000
```

**Figure 3.** Two sections of an input file for a vdW calculation. The complete input file is in the SI.

In this example, only three iterations of AutoMeKin's workflow are employed:

```
llcalcs.sh vdW.dat 100 3 48,
```

where *llcalcs.sh* is AutoMeKin's script to run all low-level calculations,[46] *vdW.dat* is the input file (summarized in Figure 3), 100 is the number of BXDE trajectories per iteration, 3 is the number of iterations, and 48 is the number of concurrent simulations in a multithreading CPU.

This calculation results in a total of 112 minima and 115 TSs for the $NO_2$ + pyrene system. Ignoring the unconnected minima, the reaction network and its corresponding minimum energy structures are displayed in Figure 4.

In sum, the more general definition of the adjacency matrix of Eq. 3 permits exploring both covalently and non-covalently bound structures, as seen in Figure 4. In particular, nine of the 12 structures of Figure 4 present covalent bonds between $NO_2$ and pyrene, while three correspond to vdW structures. The geometries of all the structures, including those not connected to the network of Figure 4 (the vast majority), are collected in the SI.

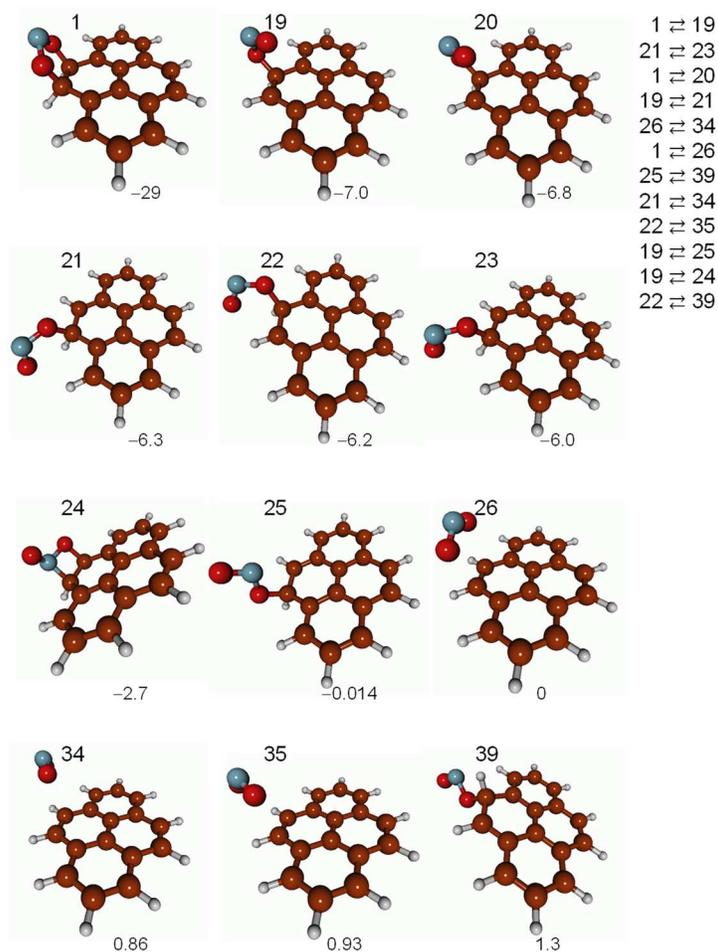

**Figure 4.** Example of a simple reaction network (using only 3 iterations of the vdW workflow; see text) obtained for the pyrene + $NO_2$ system featuring some covalent and non-covalent interactions. Numbers in the top left of each minimum energy structure are the labels, and the numbers at the bottom are their relative energies in kcal/mol.

**Chemical Knowledge (ChemKnow)**

The potential gain in efficiency of ChemKnow comes from: i) exploring only those reactions in which there is an interest, ii) imposing limits for the minimum and maximum number of neighbors (valencies) of an atom, and iii) restricting the maximum number of bonds that can break and form in each step.

The method also benefits from working in graph space, where reactions are simply graph transformations. This has been shown to be a practical way to explore reactive events by Habershon and co-workers.[26,27,97-100]

The workflow of ChemKnow is detailed as follows:

1) The reactive sites (*active* atoms) of the system are selected, along with the maximum number of bonds that can break ($n_b$) and form ($n_f$) per elementary step, the allowed minimum (*min_val*) and maximum valencies (*max_val*) of each atom, and the maximum energy (*emax*) of the system.
2) Beginning with a given minimum energy geometry turned into its graph analogue, ChemKnow generates all possible graph transformations that comply with the constraints of the previous step. Three additional restrictions are imposed on top of those specified by the user: 1) reactions where the closest distance between the (linear) paths followed by the atoms in their rearrangement is lower than a threshold value are ignored, 2) bond formations between atoms that are at a distance greater than a certain value *startd* are not allowed; and, 3) only those bonds whose bond order is lower than 1.5 can be broken.
3) The newly-generated graphs are converted back into 3D structures using constrained Langevin dynamics, with external forces applied to the active atoms. The adjacency matrix is monitored along the trajectory, and the constraints are lifted once the product graph is obtained. At this point, the final geometry is optimized, and its connectivity is checked once more to make sure that the desired product is obtained. As a second check, its energy must be lower than *emax* to retain the newly generated geometry.
4) A path connecting the initial and final geometries is constructed using the Nudged Elastic Band (NEB) method, and the highest point along the path is subjected to TS optimization.
5) Successive iterations of AutoMeKin start from a new (connected) graph, and steps 1-4 are repeated until no new graphs are found.

To avoid sampling equivalent paths multiple times in step 2, a descriptor for each unique TS is book-kept in a Python dictionary. The chosen descriptor is the list of eigenvalues of a TS adjacency matrix:

$$\mathbf{A}_{TS} = \frac{1}{2}(\mathbf{A}_R + \mathbf{A}_P) \tag{4}$$

where $\mathbf{A}_R$ and $\mathbf{A}_P$ are the reactant and product adjacency matrices, respectively, with the atomic numbers filling the diagonals. This descriptor has the property of being invariant with respect to permutations of like atoms, thus avoiding sampling equivalent paths more than once.

Additionally, steps 3 and 4 of the above pipeline are carried out using ASE's ExternalForces and AutoNEB classes, respectively,[90] with all the graph analysis and transformations performed using NetworkX.[82]

By way of example, ChemKnow was employed to study the fragmentation channels of formic acid (FA) using the following constraints:

a) All atoms are *active*.
b) $n_f \leq 2$ and $n_b \leq 2$, where transformations with both $n_f = 2$ and $n_b = 2$ are discarded.
c) *min_val* = [1,1,1] and *max_val* = [1,4,2] for the list of atoms = [H,C,O], respectively.
d) *startd* = 2.75 Å
e) *emax* = 150 kcal/mol.

When the initial structure is *cis*-FA, a total of 77 distinct $(n_f,n_b)$ combinations are found, which breaks down into 3 (0,1); 3 (0,2); 5 (1,0); 15 (1,1); 15 (1,2); 9 (2,0); and 27 (2,1) combinations. This number excludes the cleavage of bonds with bond orders greater than 1.5. Additionally, when the other constraints were imposed, the number of combinations (paths) that start in *cis*-FA became 8.

Overall, this approach only needs four iterations of the workflow to reach convergence (no further minima found) for FA, which affords a total of 8 TSs and 4 minima at the PM7 SQM level after exploring a total of 24 paths. The $CO_2$:CO branching ratio obtained at 150 kcal/mol of excitation energy is 3:97. In comparison, an MD sampling with 200 trajectories leads to 11 TSs and 7 minima and a $CO_2$:CO ratio of 2:98.

A test was also done on vinyl cyanide (VC), to compare the performance of ChemKnow *vs* an MD-based sampling. Chemknow's constraints are similar to those employed for FA, including now (2,2) combinations of $(n_f,n_b)$ for the graph transformations, and *min_val* and *max_val* for N are 1 and 4, respectively. In this example, ChemKnow needs to sample 850 paths to obtain 59 TSs and 31 minima *vs* 2000 trajectories employed by the MD module, which affords 64 TSs and 27 minima. Although the efficiency of ChemKnow is superior to MD for VC, the former failed to optimize a significant TS connecting VC to vinyl iso-cyanide.[71]

On the other hand, if the set of active atoms and range of valences are reduced, ChemKnow probably outperforms MD-based methods in terms of efficiency. However, the decision to employ this method should also rely on its efficacy in finding the relevant structures for the system under study.

It was noted in passing that, while the MD-based methods have been heavily tested, ChemKnow needs further assessment and perhaps an optimization of steps 3 and 4, which are the major components of the TS search algorithm.

**New TS search algorithm (bots)**

As an alternative to the geometry-based *bbfs* algorithm described above, a new method to detect reactive events, recently put forward by Wang and co-workers,[81] is also available in AutoMeKin. The method is called *bots*, which stands for bond-order time series. This is the workflow of *bots*:[81]

1) A low-pass filter is applied to remove the fast fluctuations from the time series using a cutoff frequency $\omega$.
2) The first derivative of the smoothed time series is obtained using the central difference formula.

3) A threshold $\mu$ is applied to the first-order derivatives to select the peaks (those above $+\mu$ and below $-\mu$).
4) As in *bbfs*, peaks within an adaptive time window[45] are merged and regarded as multi-bond reactive events.

Tests showed that this algorithm works best with BXDE because high-energy MD simulations do not typically give rise to high-frequency fluctuations in the bond-order times series. Therefore, the use of *bots* is restricted to BXDE-based methods.

Figure 5 shows part of an input file to run a BXDE sampling, followed by *bots* analysis of the trajectories. The keyword *post_proc* is employed to select *bbfs* (the default) or *bots*. In the latter case, two parameters are required: $\omega$, in cm$^{-1}$, and $\mu$, given as a multiple of $\sigma$.

```
--Method--
sampling        BXDE
ntraj           1
post_proc       bots 200 2.5
```

**Figure 5.** One section of an input file that employs *bots* TS search algorithm. The complete input file is in the SI.

By way of example, the simple test case FA is employed to run 100 BXDE trajectories, which were then analyzed using *bots* with the parameters in Figure 5. The resulting structures and kinetics are very similar to the ones obtained using the standard *bbfs* method.

Figure 6 shows the variation of three bond orders and their time derivatives for a reactive BXDE simulation leading to $H_2O + CO$. In the figure, the three bonds correspond to those that change from reactant to products.

The use of *bots* requires fine-tuning two parameters ($\omega$ and $\mu$) for the system under study. These values should be selected to find as many peaks (reactive events) as possible, while minimizing the number of false positives. Figure 6 shows that the only reactive event is successfully detected at ~3.1 ps, *i.e.*, the breaking of the 1–3 and 1–4 bonds and formation of the 3–4 bond. However, this comes at the cost of finding six false positives, at 1.4, 1.8, 1.9, 2.7, 2.8, and 2.9 ps. Although false positives also occur in *bbfs*, the major disadvantage of *bots* is its dependence on two parameters that strongly affect its performance.

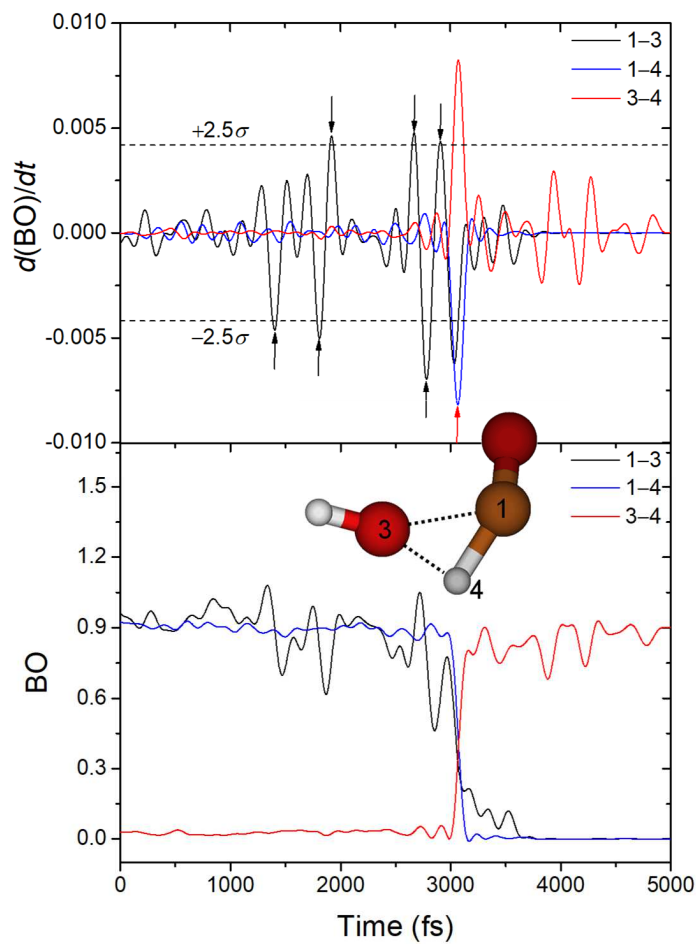

**Figure 6.** Bond orders (bottom) and their time derivative (top) for a reactive BXDE trajectory starting from *cis*-FA. The reactive event occurs at 3.1 ps and is successfully detected by *bots* (red arrow in the top panel). Black arrows correspond to false positives.

**Properties of the reaction network**

In AutoMeKin2021, properties of the reaction networks are analyzed using the NetworkX Python library.[101] Here, a node of the graph is either an intermediate or a product, while an edge represents a pathway connecting two nodes. Two types of networks are constructed in the example shown here. In the "all-states" network, every single intermediate constitutes a node, while in the "coarse-grained" one a family of conformers is lumped together to form each node. Additionally, in both networks, edges have weights representing the number of pathways connecting a pair of nodes. Finally, self-loops are avoided by removing paths connecting permutation-inversion isomers of the same node.

Common properties of a network can be studied in this new version of the program: the average shortest path length, the average clustering coefficient, the transitivity, and the assortativity. Some example systems that have been described using similar approaches are the network of organic chemistry,[102] the network involved in the ozonolysis of α-pinene,[76] and a network of small clusters.[103] Below, we give a brief description of the properties provided in AutoMeKin's output.

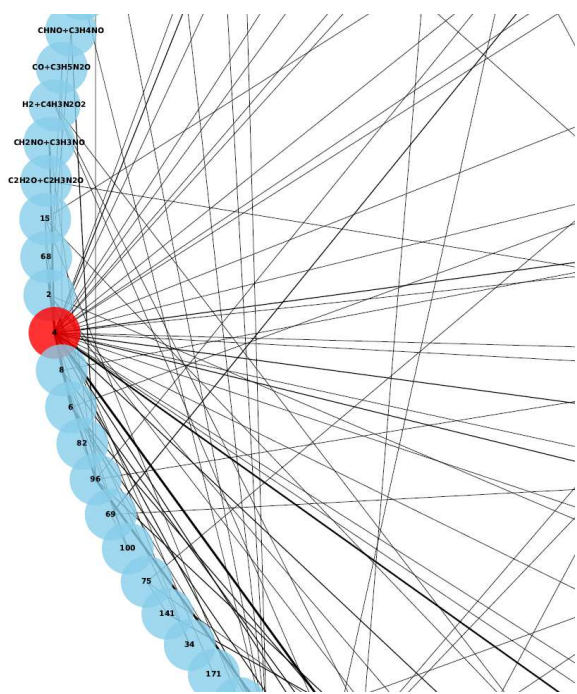

**Figure 7.** Part of the complex reaction network (in circular layout) involved in the fragmentation of protonated uracil. Nodes represent families of conformers (coarse-grained network), and the width of the edges is a measure of the number of paths between a pair of nodes. The red node corresponds to the starting structure.

The shortest path is the one connecting a pair of nodes through the least number of edges.[104] The clustering coefficient indicates the degree to which the neighbors of a node are also neighbors of each other, and an average clustering coefficient can be calculated for the network.[105] In turn, the transitivity is proportional to the ratio of the number of triangles over the number of triads in the network. For any three nodes in the network, a triangle is formed when the three possible pairs of nodes are connected, while in a triad only two pairs are connected.

The assortativity is a measure of the tendency of nodes to have connections with nodes of a similar degree and can be measured through a coefficient[106] that varies from −1 to 1. Values close to 1 indicate that nodes have a preference to connect with nodes of a similar degree, which is called assortative mixing, while values close to −1 indicate the opposite and is called disassortative mixing.

Figure 7 shows part of the coarse-grained network constructed from the results obtained with AutoMeKin for the decomposition of protonated uracil, which was chosen as an example. Table 2 shows the properties of the "all-states" and "coarse-grained" networks obtained from the same results.

| Table 2. Properties of the reaction networks[a] | | |
|---|---|---|
|  | All-states[b] | Coarse-grained[c] |
| Nodes | 208 | 116 |
| Edges | 244 | 136 |
| Density of edges (%)[d] | 1.1 | 2.0 |
| Average shortest path length | 5.49 (0.5) | 3.89 (0.4) |
| Average clustering coefficient | 0.026 (3.5) | 0.070 (5.0) |
| Transitivity | 0.019 (1.7) | 0.033 (1.7) |
| Assortativity | −0.024 | −0.18 |

[a] Numbers in parenthesis give the ratio of the value of the property over the corresponding value of a random (Erdös-Rényi) network with the same number of nodes and edges.
[b] Every structure is a node in the network.
[c] Families of conformers form a node of the network.
[d] Percentage of edges with respect to the maximum number of edges between the nodes of the network.

In general, the networks of chemical reactions are sparsely connected[76] with a low density of edges (1-2% in this example). An important feature of any network is whether they present small-world behavior, *i.e.,* when pairs of nodes are connected through a small number of edges. This property can be assessed by comparing the transitivity values and the average shortest path length with those of random networks. In this case, average shortest path lengths (5.49 and 3.89) are considerably shorter than those for the corresponding random networks, and the transitivities are 1.7 times greater. These results point out a clear "small-world" behavior. Similar results of other chemical reaction networks can be found in the literature.[76,102,103]

Clustering coefficients provide the proportion of interlinking between neighbors of a given node. The so-called scale-free networks are characterized by an enhanced clustering compared with a random network, just like the networks in this study (see Table 2).

Finally, the negative values of the assortativity indicate disassortativity mixing. That is, nodes of different degree tend to be connected. Disassortative mixing has also been observed in the ozonolysis of α-pinene,[76] in the network of organic chemistry,[102] and in biological and technological networks.[107]

The detailed reaction networks corresponding to the fragmentation of protonated uracil can be found in the SI.

**Web application**

The web application is available at https://rxnkin.usc.es/amk/ and works on most widely-used web browsers. It is for demonstration and test purposes only. Therefore, reaction mechanisms and kinetics results are predicted at the PM7 SQM level of theory and the maximum number of atoms is limited to 15.

Figure 8 shows three screenshots of the most relevant sections of the web application. Briefly, users first need to register using a valid email account. Once this is done they can request a "New Job", whose details need to be specified (Figure 8b): geometry of the system, charge, and the temperature or energy of the kinetics simulations.

To input the geometries, a JSmol viewer integrated in our web interface was employed.[108] Once all parameters are specified, users can submit their jobs by clicking the "submit your job" button. The number of jobs is not limited, but users should try to limit the number of jobs they submit at once so that other users can also use the server at the same time.

The status of the jobs is shown on a different page (Figure 8c). Depending on the size of the system and workload, the execution time can vary substantially, and users can log out. Upon job completion, they will receive a notification in their email account.

**Figure 8.** Different screenshots of the web application featuring: a) the front page, b) the area employed to set up your calculation, and c) the job queue.

Finished jobs appear with a "Completed" status in Figure 8c and users will be able to download a brief summary of the results in PDF format ("Report"), as well as a tarball file with detailed data ("Data").

The web interface was built following HTML5 recommendations.[109] The Bootstrap framework[110] is included to develop a responsive design. The Apache HTTP server, the MariaDB server, and PHP are used to build the backend. A batch system written in Perl and C deals with the execution of each job, balancing the system workload, updating the status of the jobs, moving the results to the Apache download area and notifying the users upon job completion.

**Other improvements**

This new version also includes the possibility of employing Entos Qcore[84] for the electronic structure calculations. An example input file (FA_qcore.dat) employing this option for both low-level and high-level calculations can be found in the examples folder. The corresponding test can be run using:

```
run_test.sh --tests = FA_qcore
```

Briefly, these calculations can be requested through the keywords *LowLevel* and *HighLevel*. An example of a low-level calculation using Entos Qcore is also given above for the pyrene + $NO_2$ system. For the high-level calculations, the syntax is:

```
HighLevel qcore qcore_template
```

Where *qcore_template* is the name of a file that contains the instructions to carry out the Entos Qcore high-level calculations:

```
  dft(
  xc = PBE
  ao = '6-31G*'
  )
```

Since IRC calculations are not available in Entos Qcore, a damped velocity Verlet algorithm[111] is utilized to follow the reaction pathways.

A useful feature to study the decomposition of ions is the assignment of charges (and multiplicities) of the resulting fragments. This option is only available for high-level calculations using Gaussian.[85] Charges and multiplicities are assigned using the keyword pop=(mk,nbo). This keyword is added to a single point calculation for the geometry of the last point of an IRC leading to fragmentation.

As an example, Figure 9 shows part of the high-level pathways obtained for the decomposition of protonated uracil at the B3LYP/6-31+G(d,p) level of theory. The figure displays only a reduced part of the pathways (those involving different fragments) with the positive charges assigned to the corresponding fragments.

```
TS #     DE(kcal/mol)                   Reaction path information
====     ============                   =========================
  98         60.0                  MIN    12 ---->        PR5:  CHNO + C3H4NO+
 143         72.0                  MIN   224 ---->        PR6:  C4H3N2O+ + H2O
 144         72.2                  MIN    50 ---->        PR26: C4H4N2O2+ + H
 300         98.4                  MIN    10 ---->        PR46: C3H5N2O+ + CO
 369        110.5   PR6:  C4H3N2O+ + H2O <---> PR38: C2HN2O+ + C2H2 + H2O
 396        114.0                  MIN     6 ---->        PR40: CH2NO+ + C3H3NO
 421        118.7                  MIN   216 ---->        PR29: C4H3N2O2+ + H2
 450        123.9                  MIN   467 ---->        PR22: CH2NO+ + CO + C2H3N
 497        140.4                  MIN   432 ---->        PR12: CH4NO+ + C3HNO
 504        143.4   PR8:  C4H3N2O+ + H2O <--->            PR52: C4H2N2O + H3O+
```

**Figure 9.** Pathways involving different fragments for the high-level network of protonated uracil. A comprehensive list of all pathways is given in the SI.

Since the reaction detection algorithms focus on bond formation/breakage, there is an additional tool to scan dihedral angles and find transition states for interconverting conformers. Torsions around bonds with bond orders greater than 1.5 and/or those belonging to rings are excluded.

Finally, an auto-installer script is now available, which eases the burden of installing third-party packages. The script installs singularity[112] and downloads the latest container image from sylabs (https://sylabs.io/). An instance of the container is started using a sandbox image deployed under $(TMPDIR-/tmp) folder. The container comes with all AutoMeKin's tools installed in $AMK.

## Conclusions

Presented here is the open-source software package AutoMeKin, for automated reaction discovery. AutoMeKin is an updated version of tsscds2018 featuring several new tools: rare-event MD simulations, a search algorithm to study van der Waals complexes, a chemical-knowledge based search procedure, a reactive-event detection method based on bond orders, statistics of the chemical reaction networks, and a web application to submit online jobs.

AutoMeKin is actively developed, and the most relevant functionalities that will be incorporated in the future include (but are not limited to):

a) An interface with M3C[113] to study fragmentation of vibrationally excited molecules including barrierless mechanisms.
b) A deep-learning correction to SQM barrier heights to boost the performance and efficiency of the calculations.[114]
c) An interface with Pilgrim,[115] a code to calculate thermal rate constants of chemical reactions including variational and tunneling effects.

## Acknowledgments

This work was partially supported by the Ministerio de Ciencia e Innovacion (Grant # PID2019-107307RB-I00). GLB gratefully acknowledges support from the National Science Foundation under grant No. 1763652.

**Keywords:** Reaction mechanisms, kinetics, MD simulations, Graph Theory.

# Supporting Information (SI)

## vdW

<u>Input files</u>

**vdW.dat**

```
--General--
molecule  pyrene-NO2
fragmentA pyrene
fragmentB NO2
LowLevel  qcore xtb

--Method--
sampling vdW
rotate    com com 4.0 1.5
Nassoc    10
ntraj     1
fs        2000

--Screening--
MAPEmax 0.0001
BAPEmax 0.5
eigLmax 0.01

--Kinetics--
Energy 150
```

**pyrene.xyz**

```
26

C    0.000000     1.214210    -2.812580
C    0.000000     1.223960    -1.409630
C    0.000000     0.000000    -0.704710
C    0.000000    -1.223960    -1.409630
C    0.000000     0.000000     0.704710
C    0.000000    -1.223960     1.409630
C    0.000000    -2.435000     0.699840
C    0.000000    -2.435000    -0.699840
C    0.000000     1.223960     1.409630
H    0.000000    -3.380390    -1.229240
C    0.000000    -1.214210    -2.812580
C    0.000000     0.000000    -3.507360
C    0.000000    -1.214210     2.812580
H    0.000000    -3.380390     1.229240
C    0.000000     1.214210     2.812580
C    0.000000     0.000000     3.507360
C    0.000000     2.435000     0.699840
```

```
H    0.000000   -2.144130    3.368670
H    0.000000    2.144130    3.368670
H    0.000000    0.000000    4.589720
H    0.000000    2.144130   -3.368670
C    0.000000    2.435000   -0.699840
H    0.000000   -2.144130   -3.368670
H    0.000000    0.000000   -4.589720
H    0.000000    3.380390    1.229240
H    0.000000    3.380390   -1.229240
```

## NO2.xyz

```
3

N    3.073876    0.000123   -0.318014
O    3.181202   -1.096194    0.133758
O    3.181114    1.096449    0.133758
```

**Results:** http://doi.org/10.5281/zenodo.4700262

# ChemKnow

Input files

## FA_ck.dat

```
--General—
molecule   FA
LowLevel   mopac pm7
charge     0
mult       1

--Method--
sampling ChemKnow

--Screening--
imagmin 200
MAPEmax 0.008
BAPEmax 2.5
eigLmax 0.1

--Kinetics--
Energy 150
```

## FA_md.dat

```
--General--
molecule   FA
LowLevel   mopac pm7
```

```
charge   0
mult     1

--Method--
sampling MD
ntraj    10

--Screening--
imagmin 200
MAPEmax 0.008
BAPEmax 2.5
eigLmax 0.1

--Kinetics--
Energy 150
```

**FA.xyz**

```
5

C     0.000000      0.000000      0.000000
O     0.000000      0.000000      1.220000
O     1.212436      0.000000     -0.700000
H    -0.943102      0.000000     -0.544500
H     1.038843      0.000000     -1.634005
```

Input files for VC_ck and VC_md are identical to those for FA, except for the acronym VC and the initial XYZ geometry file (VC.xyz)

**VC.xyz**

```
7

C    -0.68560     -0.52150     -0.00050
C    -1.65160      0.39340      0.00010
C     0.68710     -0.11350     -0.00010
N     1.77600      0.21020      0.00010
H    -0.93660     -1.57190      0.00330
H    -2.68680      0.08560     -0.00010
H    -1.40050      1.44380      0.00080
```

**Results:** http://doi.org/10.5281/zenodo.4700257

# New TS search algorithm (bots)

<u>Input files</u>

**FA_bots.dat**

```
--General--
molecule   FA
LowLevel   mopac pm7
charge     0
mult       1

--Method--
sampling   BXDE
ntraj      1
post_proc bots 200 2.5

--Screening--
imagmin 200
MAPEmax 0.008
BAPEmax 2.5
eigLmax 0.1

--Kinetics--
Energy 150
```

The XYZ geometry file is the same as above (for ChemKnow)

**Results**: http://doi.org/10.5281/zenodo.4700241

# Protonated uracil

<u>Input files</u>

**uracil.dat**

```
--General--
molecule   uracil
LowLevel   qcore xtb
HighLevel g09 b3lyp/6-31+G(d,p)
IRCpoints 30
charge     1
mult       1

--Method--
sampling MD
ntraj    10

--Screening--
```

```
imagmin 200
MAPEmax 0.008
BAPEmax 2.5
eigLmax 0.1

--Kinetics--
Energy 150
```

**uracil.xyz**

```
13

N 0.000000 1.016495 0.000000
C 1.112839 0.313955 0.000000
N 1.126388 -1.043152 0.000000
C -0.049649 -1.744872 0.000000
C -1.235176 -1.071681 0.000000
C -1.161437 0.348965 0.000000
O 2.301950 0.873672 0.000000
H 2.024196 -1.520859 0.000000
H 0.033025 -2.825098 0.000000
H -2.189572 -1.581794 0.000000
O -2.284621 1.026810 0.000000
H -2.116968 1.987348 0.000000
H 2.226502 1.844945 0.000000
```

**Results:** http://doi.org/10.5281/zenodo.4711089